\documentclass[aps,prd,preprint,floats,superscriptaddress,nofootinbib]{revtex4}
\usepackage{epsfig}

\begin{document}
\def \epj#1#2#3{{Eur. Phys. J. }{\bf#1}, #2 (#3)}
\def \etal{{\it et\,al.}}
\def \nima#1#2#3{{\it Nucl. Instr. Meth.} {\bf A#1} (#3) #2}
\def \np#1#2#3{{Nucl. Phys.} {\bf#1},  #2 (#3)}
\def \npa#1#2#3{{Nucl. Phys.} {\bf A#1}, #2 (#3)}
\def \npb#1#2#3{{Nucl. Phys.} {\bf B#1}, #2 (#3)}
\def \epl#1#2#3{{Euro. Phys. Lett.} {\bf#1} (#3) #2}
\def \pl#1#2#3{{Phys. Lett.} {\bf#1} (#3) #2}
\def \plb#1#2#3{{Phys. Lett. B} {\bf #1} (#3) #2}
\def \pr#1#2#3{{Phys. Rev.} {\bf#1} (#3) #2}
\def \pra#1#2#3{{Phys. Rev. A} {\bf#1}, #2 (#3)}
\def \prb#1#2#3{{Phys. Rev. B} {\bf#1}, #2 (#3)}
\def \prc#1#2#3{{Phys. Rev. C} {\bf#1}, #2 (#3)}
\def \prd#1#2#3{{Phys. Rev. D} {\bf#1}, #2 (#3)}
\def \prl#1#2#3{{Phys. Rev. Lett.} {\bf#1}, #2 (#3)}
\def \prp#1#2#3{{Phys. Rep.} {\bf#1}, (#3) #2}
\def \ptp#1#2#3{{\it Prog. Theor. Phys.} {\bf#1} (#3) #2}
\def \pht#1#2#3{{\it Phys. Today} {\bf#1} (#3) #2}
\def \rmp#1#2#3{{Rev. Mod. Phys.} {\bf#1} (#3) #2}
\def \asy#1#2{{^{+#1}_{-#2}}}
\newcommand{\dedx}{\mbox{$dE/dx$}}
\newcommand{\eb}{\mbox{$E_{{\rm beam}}$}}
\newcommand{\ebsq}{\mbox{$E^2_{{\rm beam}}$}}
\newcommand{\pip}{\mbox{$\pi$}^{+}}
\newcommand{\pim}{\mbox{$\pi^{-}$}}
\newcommand{\piz}{\mbox{$\pi$}^{0}}
\newcommand{\Bbar}{\mbox{$\overline{B}$}}
\newcommand{\Bbarzero}{\mbox{$\overline{B}^0$}}
\newcommand{\Bzero}{\mbox{${ B^0}$}}
\newcommand{\Bee}{\mbox{${ B}$}}
\newcommand{\Dee}{\mbox{${ D}$}}
\newcommand{\Kay}{\mbox{${ K}$}}
\newcommand{\Bminus}{\mbox{${ B^-}$}}
\newcommand{\Bplus}{\mbox{${ B^+}$}}
\newcommand{\BBbar}{\mbox{$B\overline{ B}$}}
\newcommand{\Kplus}{\mbox{${ K}^{+}$}}
\newcommand{\Kminus}{\mbox{${ K}^{-}$}}
\newcommand{\Dplus}{\mbox{${ D}^{+}$}}
\newcommand{\Dminus}{\mbox{${ D}^{-}$}}
\newcommand{\Dzero}{\mbox{${ D}^0$}}
\newcommand{\Dzerobar}{\mbox{$\overline{D}^0$}}
\newcommand{\Dbarzero}{\mbox{$\overline{D}^0$}}
\newcommand{\Dbar}{\mbox{$\overline{ D}$}}
\newcommand{\invfb}{\;\mbox{$fb^{-1}$}}
\newcommand{\invf}{\;\mbox{$fb^{-1}$}}
\newcommand{\goesto}{\mbox{$\rightarrow$}}
\newcommand{\ee}{\eplus\eminus}			
\newcommand{\electron}{\mbox{$e^{-}$}}			
\newcommand{\eminus}{\electron}
\newcommand{\positron}{\mbox{$e^{+}$}}
\newcommand{\eplus}{\positron}

\def \dzpi{49.7}
\def \dzpistat{\pm 1.2 }
\def \dzpisys{\pm 2.9 }
\def \dzpiferr{\pm 2.2 }

\def \dzpistatavg{50.0}
\def \dzpistatstatav{\pm 1.1 }

\def \dppi{26.8}
\def \dppistat{\pm 1.2}
\def \dppisys{\pm 2.4}
\def \dppiferr{\pm 1.2}

\def \prod{2.41}
\def \prodstat{\pm 0.11}
\def \prodsys{\pm 0.15}
\def \prodferr{\pm 0.11}

\def \cosd{0.863}
\def \cosdstat{\asy{0.024}{0.023}}
\def \cosdsys{\asy{0.036}{0.035}}
\def \cosdferr{\asy{0.038}{0.030}}

\def \cosdcleo{0.877}
\def \cosdstatcleo{\pm 0.030}
\def \cosdsyscleo{\asy{0.046}{0.044}}
\def \cosdferrcleo{\asy{0.039}{0.031}}

\def \aratio{0.69}
\def \aratiostat{\pm 0.03}
\def \aratiosys{\pm 0.06}
\def \aratioferr{\pm 0.06}

\def \deltalow{13.6}
\def \deltahigh{38.3}
\def \combdeltalow{16.5}
\def \combdeltahigh{38.1}\def \mbwid{\columnwidth}

\preprint{\tighten\vbox{\hbox{\hfil CLNS 02/1788}
                        \hbox{\hfil CLEO 02-8}
}}

\title{Measurement of 
${\cal{B}}(\Bminus\goesto \Dzero\pim)$ and
${\cal{B}}(\Bbarzero\goesto \Dplus\pim) $
and Isospin Analysis of 
$B\goesto D\pi$ Decays
}  

\author{CLEO Collaboration}
\date{\today}

\begin{abstract} 

We present
new measurements 
of branching fractions for 
the color-favored decays $\Bminus \goesto \Dzero\pim$
and $\Bbarzero\goesto \Dplus\pim$.  
Using $9.67\times 10^{6}\;\BBbar$ pairs
collected with the CLEO detector, we
obtain 
the branching fractions 
${\cal{B}}(\Bminus\goesto \Dzero\pim) = 
(\dzpi\dzpistat\dzpisys\dzpiferr) \times 10^{-4}$
and
${\cal{B}}(\Bbarzero\goesto \Dplus\pim) = 
(\dppi\dppistat\dppisys\dppiferr) \times 10^{-4}$.
The first error is statistical, 
the second is systematic, and the third is due 
to the experimental uncertainty
on the production ratio of 
charged and neutral $\Bee$ mesons
in $\Upsilon(4S)$ decays.
These results, together
with the current world average for
the color-suppressed branching fraction
${\cal{B}}(\Bbarzero\goesto \Dzero\piz)$, are used 
to determine the cosine of the strong phase
difference $\delta_I$ between the $I=1/2$ and $I=3/2$ isospin amplitudes.
We find $\cos\delta_I = \cosd \cosdstat \cosdsys \cosdferr$,
and obtain a $90\%$ confidence interval 
of $\combdeltalow^\circ < \delta_I < \combdeltahigh^\circ$.  
This non-zero value of $\delta_I$ 
suggests the presence of final state interactions in 
the $\Dee\pi$ system.
\end{abstract}
\maketitle

\newpage

{
\renewcommand{\thefootnote}{\fnsymbol{footnote}}

\begin{center}
S.~Ahmed,$^{1}$ M.~S.~Alam,$^{1}$ L.~Jian,$^{1}$ M.~Saleem,$^{1}$
F.~Wappler,$^{1}$
E.~Eckhart,$^{2}$ K.~K.~Gan,$^{2}$ C.~Gwon,$^{2}$ T.~Hart,$^{2}$
K.~Honscheid,$^{2}$
% ...
\end{center}
 
\small
\begin{center}
$^{1}${State University of New York at Albany, Albany, New York 12222}\\
$^{2}${Ohio State University, Columbus, Ohio 43210}\\
% ...
\end{center}
 
\setcounter{footnote}{0}
}
\newpage
\def \mbwid{\columnwidth}

This paper presents the results of measurements of 
the branching fractions for $\Bminus\goesto\Dzero\pim$
and $\Bbarzero\goesto\Dplus\pim$ and
the extraction of the strong phase difference $\delta_I$ 
between the 
$I = 1/2$ and $I = 3/2$ isospin amplitudes in the $\Dee\pi$ system.
These decays are an excellent 
testing ground for the 
theoretical description of hadronic $B$-meson 
decays. Our understanding of these decays has
improved considerably during the past few years with the
development and application of 
Heavy Quark Effective Theory (HQET) \cite{neubert_beneke,heavy_flavors}
and Soft Collinear Effective Theory (SCET) \cite{scet_btodpi}. 
Originating from the simple, 
but very effective, idea of color-transparency~\cite{colortrans}, 
the factorization hypothesis
has been put on a more solid basis, and in the case of 
$\Bbar\goesto\Dee\pi$, has been 
proven within the framework of SCET.

The recent observation~\cite{cleo_btod0pi0,belle_btod0pi0} 
of the color-suppressed  
$\Bbarzero \goesto \Dzero\piz$
decay\footnote{Throughout this paper, charge conjugation is implied.} 
completed the 
measurement of the $\Dee\pi$ final states and 
was used to determine the cosine of the strong phase difference 
$\cos \delta_I =0.89 \pm 0.08$, a value which is 
consistent with one.  
A value of 
$\cos\delta_I$ inconsistent
with one would signal the presence of 
final-state interactions in the 
$\Bbar\goesto\Dee\pi$ process~\cite{rosner,neubert_petrov}.
In this paper, we present improved measurements of 
branching fractions for the color-favored decays 
$\Bminus \goesto \Dzero\pim$
and $\Bbarzero\goesto \Dplus\pim$
based on a larger data set than that from which 
the previous results were obtained,
as well as a new evaluation of 
$\cos\delta_I$ which takes
into account the correlations among the various 
contributions to the overall systematic error.

This analysis uses $\ee$ annihilation data recorded with the CLEO
detector at the Cornell Electron Storage Ring.
The integrated luminosity of the data sample is 9.15~$\invf$ collected 
on the $\Upsilon(4S)$ 
(on-resonance), corresponding to $9.67\times 10^{6}\; \BBbar$ pairs,
and 4.35~$\invf$ collected 60 MeV below the $\BBbar$ threshold
(off-resonance), which is used for background studies.
The results we present in this paper 
for ${\cal{B}}(\Bminus \goesto \Dzero\pim)$
and ${\cal{B}}(\Bbarzero\goesto \Dplus\pim)$
supersede those in the CLEO publication, Ref.~\cite{alam},
which were based on a 1.3$\invf$ subset of the data used
in the present analysis.
Data were recorded with two detector configurations, CLEO II~\cite{cleo2_detector} and CLEO 
II.V~\cite{cleo2.5_detector}.
Cylindrical drift chambers in a 1.5~T 
solenoidal magnetic field measure momentum
and specific ionization $(\dedx)$ of charged particles. Photons are
detected using a CsI(Tl) crystal electromagnetic calorimeter,
consisting of a barrel-shaped central part of 6144 crystals and
1656 crystals in the forward regions of the detector (endcaps).
In the CLEO II.V configuration, the innermost tracking chamber was replaced by a
three-layer, double-sided silicon microvertex detector, and the
main drift chamber gas was changed from argon-ethane to a 
helium-propane mixture. 



In our analysis, we impose 
quality requirements on 
charged particle tracks and improve the 
purity of pion and kaon used to reconstruct $D$ mesons
by using $\dedx$ information
if the particle momentum is less than $800$ MeV$/c$. 
The neutral $D$ mesons are reconstructed using three decay modes:
$\Kminus \pip$, 
$\Kminus\pip\piz$ 
and $\Kminus\pip\pim\pip$.  
Charged $D$ mesons are similarly reconstructed 
via the mode $\Kminus\pip\pip$.
In each case, $D$ meson candidates are required to have a mass
within $3\sigma$ (standard deviations) of the 
PDG $\Dee$ mass~\cite{pdg} 
before kinematic fitting.  Resolutions for the various 
$D$ modes range from 6 to 12 MeV.


Each $\Bee$ meson candidate is reconstructed using 
the four-momentum of the mass-constrained $\Dee$ meson 
and an additional charged track in the event (assumed to be a pion).  
Candidates are then identified using the 
beam-constrained mass $M_B = \sqrt{\ebsq - p_B^2}$,
where $\eb$ denotes the beam energy and
$p_B$ the candidate momentum, 
and the energy difference $\Delta E$ defined
by $\Delta E\equiv E_D + E_\pi - \eb$,
where $E_D$ and $E_\pi$ are the $D$ meson
and $\pi$ energies, respectively.
Preselection of $B$ candidates 
requires
$M_B > 5.24$~GeV/$c^2$ and $\Delta E$ to be between 
$-50$ and 50 MeV. 
Additionally, 
we calculate the sphericity vectors~\cite{sphericity} of the $B$
daughter particles and of the rest of the event. 
We require the absolute value of the cosine of the angle between these 
two vectors to be less than $0.8$.
The distribution of this angle is strongly peaked at $\pm$1 for
continuum background and is nearly flat for $\BBbar$ events.
We also require events to satisfy $R_{2} < 0.45$, where 
$R_2$ is the ratio of the second to zeroth Fox-Wolfram moments 
of the event~\cite{fox}.
Finally, for events with more than one 
$B$ meson candidate,
the candidate with the smallest $|\Delta E|$ is chosen.

\begin{figure}[t]
\begin{center}
\includegraphics[width=\mbwid]{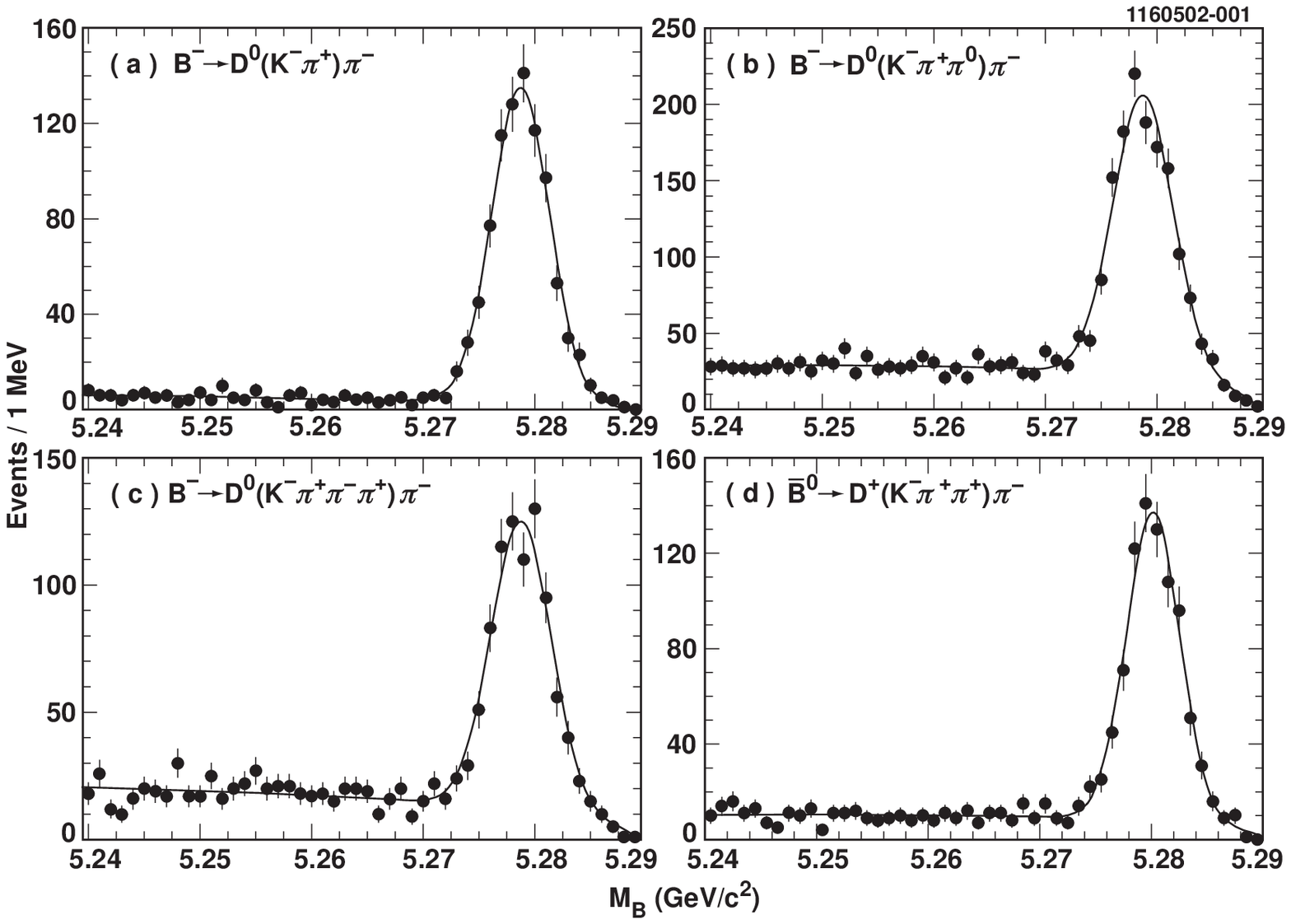}
\end{center}
\caption{
Fitted $M_B$ distributions in
for a) $\Bminus\goesto \Dzero(\Kminus\pip) \pim$
b) $\Bminus\goesto \Dzero(\Kminus\pip\piz)\pim$
c) $\Bminus\goesto \Dzero(\Kminus\pip\pim\pip)\pim$
and
d)$\Bbarzero\goesto \Dplus(\Kminus\pip\pip)\pim$.\label{mbfit}}
\end{figure}

To obtain event yields for $\Bbar\goesto\Dee\pim$ 
for each $D$ meson decay mode, the $M_B$ distribution of candidates 
surviving the
above slection cuts are fit using a 
binned maximum likelihood fit.  The function used is
a Gaussian for the signal plus an
empirical background function,
$f(M_B) = A M_B \sqrt{1-(M_B/\eb)^{2}} 
\exp{a(1-(M_B/\eb)^{2})}$,
having a fixed $\eb = 5.29$ GeV.  All other parameters in both
background and signal functions are allowed
to float in the fit.
The fitted $M_B$ distributions for each of the $D$ meson decay modes
are presented in Fig.~\ref{mbfit}.  
\begin{table*}
\begin{center}
\caption{Results for the 
branching fractions 
${\cal{B}}(\Bminus\goesto \Dzero\pim)$ 
and 
${\cal{B}}(\Bbarzero\goesto \Dplus\pim)$.  
Fit yields with errors and 
efficiencies are obtained as described in the text.
\label{dataresults}  
The errors given for the efficiencies 
correspond to the Monte Carlo statistical errors for each mode.  
The $D$ mode branching fractions and the branching fraction errors
have been taken from the 
PDG~\cite{pdg}.  
The errors reported for the measured $B$ 
branching fraction are the statistical errors only.
The current PDG average values for the two
branching fractions have been included for comparison.
}
\begin{tabular}{|l|ccc|c|}
\hline
\multicolumn{5}{|c|}{$\Bminus\goesto\Dzero\pim$}\\
\hline
$\Dzero$ Decay Mode & Yield & \rule{1mm}{0mm}Efficiency $(\%)$\rule{1mm}{0mm} & \rule{1mm}{0mm}$\Dzero$ mode ${\cal{B}} (\%)$\rule{1mm}{0mm} 
& \rule{1mm}{0mm}${\cal{B}}(\Bminus\goesto\Dzero\pim) (\times 10^{-3})$\rule{1mm}{0mm}\\
\hline
$\Kminus\pip$ & $ 820\pm 31$ & $\;45.4\pm 0.3\;$ & $3.83\pm 0.09$ & $4.90\pm 0.18$\\
$\Kminus\pip\piz$ & \rule{1mm}{0mm}$1200\pm 45$\rule{1mm}{0mm} & $17.1\pm 0.2$ & $13.9 \pm 0.9$ & $5.20\pm 0.19$\\ 
$\Kminus\pip\pim\pip$ & $740\pm 33$ & $20.9\pm 0.3$ & $7.49 \pm 0.31$ & $4.91\pm 0.22$\\
\hline
PDG & & & &$5.3 \pm 0.5$\\
\hline
\hline
\multicolumn{5}{|c|}{$\Bbarzero\goesto \Dplus\pim$}\\
\hline
$\Dplus$ Decay Mode & Yield & Efficiency $(\%)$ & $\Dplus$ mode ${\cal{B}} (\%)$ 
& ${\cal{B}}(\Bbarzero\goesto\Dplus\pim) (\times 10^{-3})$\\
\hline
$\Kminus\pip\pip$ & $764\pm 33$ & $32.8\pm 0.4$ & $9.0 \pm 0.6$ &$2.68 \pm 0.12$\\
\hline
PDG & & & &$3.0 \pm 0.4$\\
\hline
\end{tabular}
\end{center}
\end{table*}

A small, non-negligible background 
from the decay $\Bbar\goesto\Dee\Kminus$ 
contributes to the yields obtained by the fit procedure
described above.
We have, therefore, simulated this background via
Monte Carlo to determine 
the fraction of feed-through to the $\Dee\pim$ sample,
and performed a subtraction using the average of the
two 
measurements~\cite{cleodk,belledk} of
${\cal{B}}(\Bminus\goesto \Dzero \Kminus)/
{\cal{B}}(\Bminus\goesto \Dzero \pim) =  0.071\pm 0.009$ 
and the recent measurement~\cite{belledk} of
${\cal{B}}(\Bbarzero\goesto \Dplus \Kminus)/
{\cal{B}}(\Bbarzero\goesto \Dplus \pim) =  0.068\pm 0.017$.
The amount of $DK$ feed-through is found to be 
approximately $(4\pm 1)\%$ of the $D\pi$ yield.  
We then reduce the event yields obtained in the fit to the
data by this fraction.



Using 
efficiencies determined
by applying the above method of
analysis to samples of signal Monte Carlo events, we
obtain 
the branching fractions for the processes under 
investigation from the event yields corrected for
the $DK$ feed-through:
\begin{equation}
{\cal{B}}(\Bbar\goesto D\pi) =
\frac{\mbox{Corrected Yield}}
{\epsilon \times {\cal{B}}(\Dee\goesto f.s.)
\times N(\Upsilon(4S)) \times 
2f}, 
\end{equation}
where $f$ represents $f_{+-}$ or $f_{00}$, 
the charged or neutral $B$ meson production
ratios at the $\Upsilon(4S)$, as appropriate.
The corrected yields, efficiencies and 
final branching fraction obtained for each $\Dee$ decay
mode are shown in Table~\ref{dataresults}.
We have assumed $f_{+-} = f_{00} = 0.5$.

The three $\Bbar\goesto\Dee\pi$ decay 
branching fractions (the two color-favored modes, which we report
measurements for in the present paper, as well as the color-suppressed
mode $\Bbarzero\goesto\Dzero\piz$)
form a complete set of branching fractions
with which we may calculate $\cos{\delta_I}$, the cosine of the 
strong phase angle difference 
between the two isospin amplitudes $I = 1/2$ and $I= 3/2$ 
that contribute to the decay process.  
The expression for $\cos{\delta_I}$,
following Ref.~\cite{rosner}, is:
\begin{equation}
\cos{\delta_I} = 
\frac{3\Gamma(\Dminus\pip)+\Gamma(\Dbarzero\pip)-6\Gamma(\Dbarzero\piz)}
{4|{\cal{A}}_{1/2}{\cal{A}}_{3/2}|},
\label{coseqn}
\end{equation}
where the isospin amplitudes ${\cal{A}}_{3/2}$ and ${\cal{A}}_{1/2}$
are given by
\begin{eqnarray}
|{\cal{A}}_{3/2}|^{2} = \Gamma(\Dbarzero\pip),&\;{\mbox{and}}\\
|{\cal{A}}_{1/2}|^{2} = \frac{3}{2}
\left(\Gamma(\Dminus\pip)+\Gamma(\Dbarzero\piz)\right)&
-\frac{1}{2}\Gamma(\Dbarzero\pip).
\end{eqnarray}

The calculation of 
$\cos{\delta_I}$ in the 
$D\pi$ system takes into account
correlations of systematic errors between 
the two color-favored decay modes 
$\Bminus\goesto\Dzero\pim$ and $\Bbarzero\goesto\Dplus\pim$.
It also considers the fact that 
some of the systematic errors in the 
measurement of ${\cal{B}}(\Bminus\goesto \Dzero\pim)$
using the three $\Dzero$ decay modes
are correlated. 
Further, 
apart from the errors on  $f_{00}$ and $f_{+-}$ (which are anticorrelated),
we treat the errors between 
the two color-favored $\Bbar\goesto\Dee\pi$ modes and 
the color-suppressed $\Bbarzero\goesto\Dzero\piz$ mode as uncorrelated.
This treatment is justified since the systematic error on the 
color-suppressed mode is dominated by the background parameterization and 
fit uncertainties, whereas such contributions are not dominant 
for the color-favored modes~\cite{cleo_btod0pi0,belle_btod0pi0}.

We estimate the following systematic error contributions to our results for
these measurements:
$1\%$ per track for 
track finding and fitting,
$2\%$ for the total number of $\BBbar$ pairs,
$2\%$ per track for which $\dedx$ is used,
$2.5\%$ for the cuts used in the analysis
and $1\%$ for the $DK$ feed-through subtraction.
Other systematic errors include 
$2\%$ for $\piz$ finding in the case of 
the $D\goesto\Kminus\pip\piz$ submode,
$2.3\%-7\%$ for $D$-meson branching fractions,
$2-3\%$ for background parameterization and fitting,
and $0.7\%-2\%$ for Monte Carlo statistics.
The experimental errors of $4.5\%$ on the individual quantities $f_{+-}$ and
$f_{00}$~\cite{sylvia} are reported as 
a separate systematic error in our final result.

\begin{figure}[t]
\begin{center}
\includegraphics[width=\mbwid]{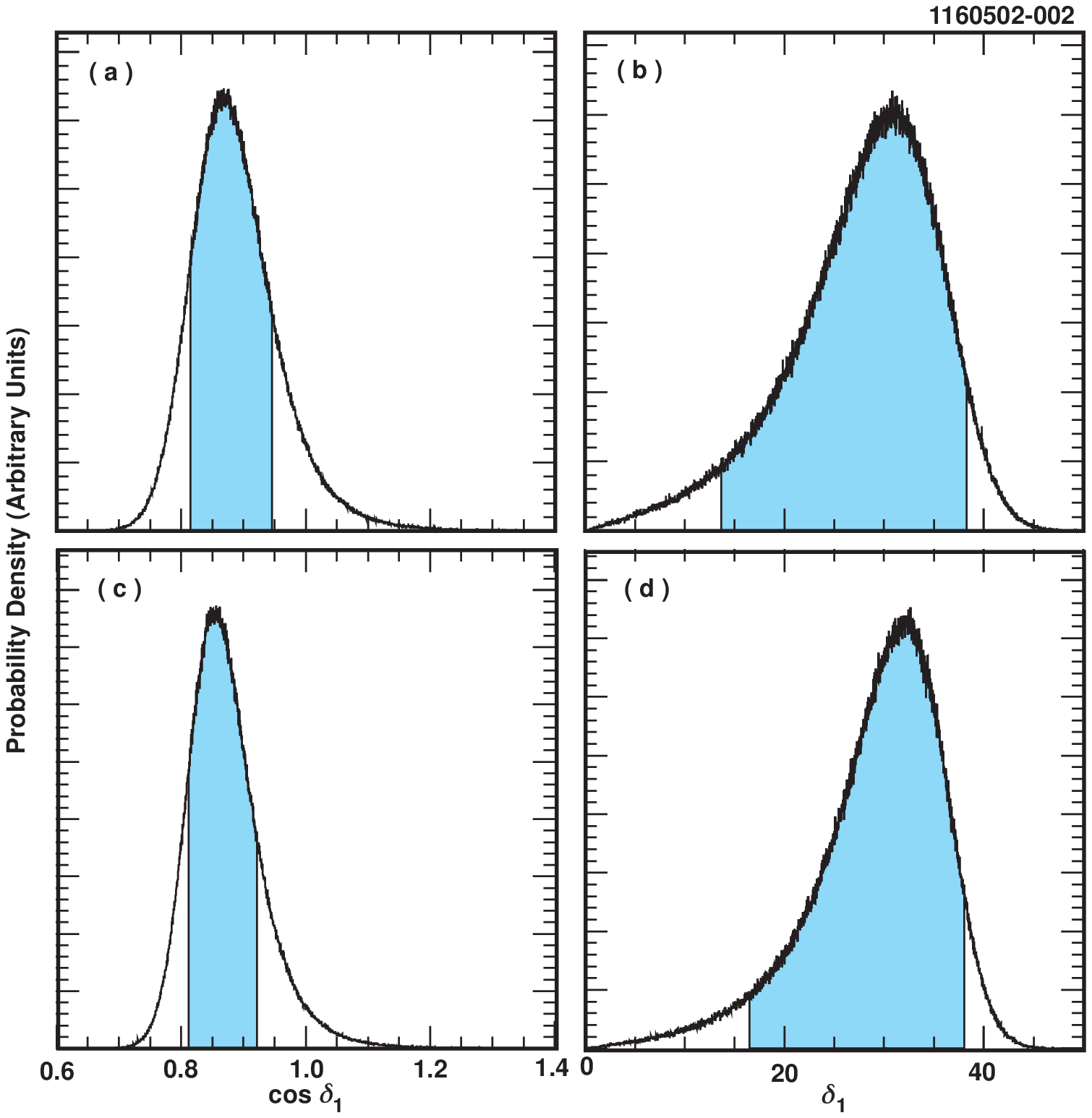}
\caption{\label{cos_dl_all.eps}The error distributions for $\cos \delta_I$ and $\delta_I$ 
obtained from the ensemble of $2\times 10^{6}$ Monte Carlo experiments
described in the text. 
The shaded area in the $\cos \delta_I$ plots is the
$\pm 1\sigma$ window (the 90\% C.L. region in the $\delta_I$ plots).
The upper two plots show the
distributions 
for a) $\cos\delta_I$ and b) $\delta_I$ 
obtained using only the CLEO measurement of 
${\cal{B}}(\Bbarzero\goesto\Dzero\pi^0)$.
The lower two plots are distributions 
for c) $\cos\delta_I$ and d) $\delta_I$
obtained using both CLEO and Belle 
measurements of
${\cal{B}}(\Bbarzero\goesto\Dzero\pi^0)$.
} 
\end{center}
\end{figure}

The overall systematic error for our measurement of 
${\cal{B}}(\Bbarzero\goesto \Dplus\pim)$ is obtained by
standard error propagation of the individual contributions.
However, in order to extract the correct overall systematic errors
for ${\cal{B}}(\Bminus\goesto\Dzero\pim)$ and 
for $\cos{\delta_I}$, we must take into 
account the correlation among the
systematic error contributions for each of the $\Dee$ submodes.
To do this, we perform Monte Carlo experiments in which we vary the measured
branching fractions by their various systematic errors.
In each experiment and for each systematic error contribution,
we generate multiplicative correction factors 
according to a Gaussian distribution. 
The combined $\Bminus\goesto\Dzero\pim$ branching fraction and 
$\cos{\delta_I}$
are then calculated from the values which have been varied as 
described above
for each Monte Carlo experiment.
From the complete ensemble of $2\times 10^{6}$ Monte Carlo experiments, 
we obtain the probability distribution functions and errors for
${\cal{B}}(\Bminus\goesto\Dzero\pim)$ and 
$\cos{\delta_I}$, which are shown in Fig.~\ref{cos_dl_all.eps}.

We thus obtain the following final results 
for the color-favored branching fractions.
\begin{eqnarray*}
{\cal{B}}(\Bminus\goesto\Dzero\pim) &=& (\dzpi\dzpistat\dzpisys\dzpiferr) \times 10^{-4},\\
{\cal{B}}(\Bbarzero\goesto \Dplus\pim) &=& 
(\dppi\dppistat\dppisys\dppiferr)\times 10^{-4}
\end{eqnarray*}
In each measured quantity, the first error is statistical 
and the second is systematic.
The third error is a
separate systematic error which corresponds to the experimental uncertainty
of the production fraction of charged (or neutral, as appropriate) $\Bee$ mesons in $\Upsilon(4S)$ 
decays.  

Our
results for ${\cal{B}}(\Bbarzero\goesto \Dplus\pim)$ 
and for ${\cal{B}}(\Bminus\goesto\Dzero\pim)$ each 
reflect improvement with respect to the 
present PDG average values~\cite{pdg}.
Our result for ${\cal{B}}(\Bbarzero\goesto \Dplus\pim)$ 
may be directly compared with the prediction of Ref.~\cite{neubert_beneke}
for this decay.  Their prediction of $32.7 \times 10^{-4}$ is marginally
consistent with our result.

The largest contribution to the overall 
systematic error in our result for 
${\cal{B}}(\Bbarzero\goesto \Dplus\pim)$ is the
$6.6\%$ relative systematic error due to the $D$ branching fraction.
We therefore report the following result which is independent of
the $\Dplus\goesto \Kminus\pip\pip$ branching fraction:
${\cal{B}}(\Bbarzero\goesto \Dplus\pim)\times 
{\cal{B}}(\Dplus \goesto \Kminus\pip\pip) = 
(\prod\prodstat\prodsys\prodferr)\times 10^{-4}.
$
 
Using the 
CLEO measurement~\cite{cleo_btod0pi0} of the color-suppressed 
branching fraction,
${\cal{B}}(\Bbarzero\goesto \Dzero\piz) = 2.74 ^{+0.36}_{-0.32}
\pm0.55 \times 10^{-4}$,
and the PDG(2002) ratio of $\Bee$ lifetimes,
$\tau(\Bplus)/\tau(\Bzero) = 1.083 \pm 0.017$,
we obtain
\[
\cos\delta_I = \cosdcleo \cosdstatcleo \cosdsyscleo \cosdferrcleo. 
\]

The error distributions derived from the ensemble of 
Monte Carlo experiments for $\cos \delta_I$ and $\delta_I$ are shown in 
Fig.~\ref{cos_dl_all.eps}. 
Integrating the $\delta_I$ distribution over the physical 
region $|\cos\delta_I| \leq 1$,
we obtain a 90$\%$ confidence interval:
\[
\deltalow^\circ < \delta_I < \deltahigh^\circ.
\]

Our final results for $\cos\delta_I$ and $\delta_I$ are based on 
the average of both measurements of
${\cal{B}}(\Bbarzero\goesto\Dzero\pi^0) = 2.92 \pm 0.45 \times 
10^{-4}$~\cite{cleo_btod0pi0,belle_btod0pi0}.
Using this average, we obtain 
\[
\cos\delta_I = \cosd\cosdstat\cosdsys\cosdferr.
\]
Similarly, we obtain our final result for 
$\delta_I$, a 90$\%$ confidence interval of
\[
\combdeltalow^\circ < \delta_I < \combdeltahigh^\circ.
\]

Using our results 
for ${\cal{B}}(\Bminus\goesto\Dzero\pim)$ and 
${\cal{B}}(\Bbarzero\goesto\Dplus\pim)$, 
we also calculate the ratio 
of the $I=1/2$ and $I=3/2$ 
isospin amplitudes,
${\cal{A}}_{1/2}/{\cal{A}}_{3/2}=
\aratio\aratiostat\aratiosys\aratioferr.$
In the heavy quark limit, ${\cal{A}}_{1/2}/{\cal{A}}_{3/2} = 1$.\footnote{The ratio given
here is based on the formalism of Ref.\cite{rosner}.  It is 
equivalent to ${\cal{A}}_{1/2}/(\sqrt{2}{\cal{A}}_{3/2})$ according 
to that of Ref.~\cite{neubert_petrov}} Corrections
to this are ${\cal{O}}(\Lambda_{\rm QCD}/m_c)$, which is consistent
with our result.  




In summary, we have measured the branching ratios for 
the color-favored 
$B\goesto D\pi$ decays, and used these measurements,
together with the current average of measurements of 
${\cal{B}}(\Bbarzero\goesto\Dzero\piz)$,
to determine the value of the cosine of the 
strong phase difference $\delta_I$ in the $D\pi$ 
system, and the ratio of $I=3/2$ and $I=1/2$ isospin amplitudes.
Our result for $\cos\delta_I$ 
differs from one by approximately
2.3$\sigma$ and thus suggests the
 presence of final-state interactions in $\Bbar\goesto\Dee\pi$ decays.


We thank Jonathan Rosner and Matthias Neubert for 
useful discussions.  
We gratefully acknowledge the efforts of the CESR staff in providing us with
excellent luminosity and running conditions.
M. Selen thanks the PFF program of the NSF and the Research Corporation, 
and A.H. Mahmood thanks the Texas Advanced Research Program.
This work was supported by the National Science Foundation, and the
U.S. Department of Energy.


\end{document}